\documentclass{aa}
\usepackage{graphicx} 

\begin{document}
\title{Reduction of chemical networks}

   \subtitle{II. Analysis of the fractional ionisation in protoplanetary discs}

   \author{D. Semenov\inst{1}, D. Wiebe\inst{2},
           \and  Th. Henning\inst{1}
           }

   \offprints{D. Semenov, \email{semenov@mpia-hd.mpg.de}}

        \institute{%
                   Max Planck Institute for Astronomy, 
                   K\"onigstuhl 17, 69117 Heidelberg, Germany\\
                   \email{semenov@mpia-hd.mpg.de, henning@mpia.de}
                \and
Institute of Astronomy of the RAS, 
                   Pyatnitskaya St. 48, 119017 Moscow, Russia\\
                   \email{dwiebe@inasan.ru}
                  }
       
   \date{Received 29 July 2003 / Accepted 21 November 2003}

\abstract{
We analyse the evolution of the fractional ionisation in a steady-state
protoplanetary disc over $10^6$~yr. We consider a disc model with a
vertical temperature gradient and with gas-grain chemistry including
surface reactions. The ionisation due to stellar X-rays, stellar and
interstellar UV radiation, cosmic rays and radionuclide decay is taken
into account. Using our reduction schemes as a tool for the analysis,
we isolate small sets of chemical reactions that reproduce the evolution
of the ionisation degree at  representative disc locations with an
accuracy of 50\%--100\%. On the basis of fractional ionisation, the disc
can be divided into three distinct layers. In the dark dense midplane
the ionisation degree is sustained by cosmic rays and radionuclides
only and is very low, $\la 10^{-12}$. This region corresponds to the
so-called ``dead zone'' in terms of the angular momentum transport
driven by MHD turbulence. The ionisation degree can be accurately
reproduced by chemical networks with about 10 species and a similar
number of reactions. In the intermediate layer the chemistry of the
fractional ionisation is driven mainly by the attenuated stellar X-rays
and is far more complicated. For the first time, we argue that surface
hydrogenation of long carbon chains can be of crucial importance for the
evolution of the ionisation degree in protoplanetary discs. In the
intermediate layer reduced networks contain more than a 100 species and
hundreds of reactions. Finally, in the unshielded low-density surface
layer of the disc the chemical life cycle of the ionisation degree comprises
a restricted set of photoionisation-recombination processes. It is
sufficient to keep about 20 species and reactions in reduced networks.
Furthermore, column densities of key molecules are calculated and compared
to the results of other recent studies and observational data. The relevance
of our results to the MHD modelling of protoplanetary discs is discussed.

        }
\titlerunning{Reduction of chemical networks. II}
\authorrunning{D. Semenov et al.}

   \maketitle

\section{Introduction}

Nowadays a paradigm of the evolution of a protoplanetary disc is widely accepted in
which most of the disc matter is assumed to move steadily toward a protostar due to 
redistribution of the angular momentum. Ionisation in such objects is an important factor that
enables the angular momentum transport to occur via magnetohydrodynamic (MHD) turbulence driven by the
magnetorotational instability (MRI; Balbus \& Hawley~\cite{MRI}). From this point of view, a disc is
conventionally divided into the ``active'' layer, adjacent to the disc surface, and the
``dead'' zone, centred on the midplane. The active part of the disc is irradiated by
high energy stellar/interstellar photons. Thus, the fractional ionisation there is relatively high, which
implies that the magnetic field is well coupled to the gas. Due to this coupling, the active layer is 
unstable to the MRI, and the developing turbulence allows the accretion to occur (e.g., Gammie~\cite{gammie}; 
Fleming \& Stone~\cite{fs2003}). The shielded midplane region is almost neutral, decoupled from the 
magnetic field, and, thus, quiescent.

The location of the boundary between these two regions may prove to be very sensitive
to the disc physical properties and chemical composition (e.g., Fromang, Terquem,
\& Balbus~\cite{from2002}). However, the MHD modelling with non-ideal effects included
is a very demanding computational task. This is why in the MHD
modelling of protoplanetary discs (and protostellar clouds) a very simple chemical scheme is usually assumed with a few 
ions and a network that includes only ionisation and recombination reactions, often neglecting the presence
of dust grains (Sano et al.~\cite{sano}; Fromang et al.~\cite{from2002}; Fleming \& Stone~\cite{fs2003}). 
The medium is believed to be in chemical equilibrium so that the fractional ionisation $x_\mathrm{e}$ can be 
expressed as (e.g., Gammie~\cite{gammie})
\begin{equation}
x_\mathrm{e}(\mathrm{eq})=\sqrt{\zeta\over\beta n_\mathrm{H}},
\label{xe}
\end{equation}
where $\zeta$ is the ionisation rate, $\beta$ is a typical recombination coefficient,
and $n_\mathrm{H}$ is the hydrogen number density. This approach may indeed be valid if only the
cosmic ray ionisation is taken into account and only gas-phase chemical processes are considered. 
However, dust plays an important role in the evolution of the fractional ionisation
being an efficient electron donor for recombining ions and a sink for neutrals in cold parts of 
a disc. Moreover, newly born stars possess a relatively high X-ray flux with photon energies
from about 1 to 5 keV (e.g. Igea \& Glassgold~\cite{IG99}). These factors give rise to a more 
complicated chemistry relevant for the ionisation degree. In particular, different ions dominate
the fractional ionisation at different times and in different parts of the disc. Therefore,
the straightforward application of Eq.~(\ref{xe}) for evaluating the ionisation degree can be 
fraught with errors in some parts of a disc.

To check the validity of Eq.~(\ref{xe}), we analyse the detailed ionisation
structure of a protoplanetary disc computed with the full UMIST\,95 chemical
network (Millar et al.~\cite{umist95}) with the surface chemistry included.
We adopt a reference disc model to serve as a guide to the range of possible
physical conditions that may be encountered in real protoplanetary objects.
Within this model, we choose several representative disc locations and
investigate in detail which chemical processes control the time-dependent
ionisation degree there. Our intention is to show that the oversimplified
treatment of the ionisation in a protoplanetary disc as an equilibrated
ionisation-recombination cycle can lead to $x_\mathrm{e}$ values that differ
by more than an order of magnitude from values computed 
with all the available information on the disc chemical evolution.

We describe the chemical processes that are responsible for ionisation in a
disc in terms of reduced networks that are subsets of the full UMIST\,95
network, supplied in some cases by a few surface reactions. These networks
contain only those species and reactions that are needed to reproduce
$x_{\rm e}$ with up to a factor of 2 uncertainty. The utilised reduction
methods are described in Wiebe, Semenov \& Henning (\cite{papi}, hereafter
Paper~I). The species-based reduction rests upon the Ruffle et
al.~(\cite{Rea02}) technique and consists of choosing species that are
important in a particular context, and then selecting from the entire network only those 
species that are necessary to compute abundances of important species with 
a reasonable accuracy. In the reaction-based method, analysis starts
from reactions that govern the 
abundance of important species. All reactions in the entire network are 
assigned weights according to the influence they have on abundances of
important species. Then, only those reactions are selected that have weights 
above a cut-off parameter that is selected on the basis of the requested 
accuracy. Only those species are included in the reduced network that 
participate in selected reactions. 

The organisation of the paper is the following. In Section~\ref{mod} we
describe the disc model and its physical structure as well as updates
to the chemical model of Paper~I.
In Section~\ref{icon} we discuss the initial conditions for the disc chemistry.
The processes responsible for the fractional ionisation in various parts of
the disc are outlined in Section~\ref{res}. In Section~\ref{colden} column densities are tabulated and 
compared to other studies and observational data. Results of the analysis and their
relevance to the MHD modelling are discussed in
Section~\ref{diss}. Final conclusions are drawn in Section~\ref{concl}.

\section{Disc model and physics update}
\label{mod}

\subsection{Disc structure and parameters}
\label{dstruc}

As a basis for our calculations, we adopted the steady-state disc model of D'Alessio et al.~(\cite{DALdisc}). 
The central star is assumed to be a classical T Tau 
star with an effective temperature $T_*=4\,000$~K, mass $M_*=0.5M_{\odot}$, and radius $R_*=2R_{\odot}$.
The disc has a radius of 373~AU, an accretion rate $\dot{M}=10^{-7}\,M_\odot$~yr$^{-1}$, and viscosity 
parameter $\alpha=0.01$. It is illuminated by UV radiation from a central star and by 
interstellar UV radiation. The intensity of the stellar UV field is described with the standard $G$ factor, 
which is scaled in a way that the unattenuated stellar UV field at $R=100$~AU is a factor of $10^4$ stronger 
than the mean interstellar radiation field (Draine~\cite{Draine}). The visual extinction of stellar light at a given disc point is calculated as
\begin{equation}
A_\mathrm{V}={N_\mathrm{H}\over1.59\cdot10^{21}}~\frac{\mathrm{mag}}{\mathrm{cm}^{-2}},
\end{equation}
where $N_\mathrm{H}$ is the column density of hydrogen nuclei between the point and the star. 
The extinction of the interstellar UV radiation is computed in a similar way (but in a vertical direction only).

In addition to the UV field, three other ionisation sources are considered: 
cosmic ray ionisation ($\zeta_\mathrm{CR}$), decay of radionuclides ($\zeta_\mathrm{R}$), and 
stellar X-ray ionisation ($\zeta_\mathrm{X}$).
The following expression is used to compute the cosmic ray ionisation rate:
\begin{equation}
\begin{array}{l}
\zeta_\mathrm{CR}=\\
\quad\frac12\zeta_0\left[\exp\left({-\Sigma_1(z,r)/10^2}\right)+\exp\left({-\Sigma_2(z,r)/10^2}\right)\right],\\
\end{array}
\end{equation}
where $\Sigma_1(z,r)$ is the surface density (g~cm$^{-2}$) {\em above} the point
with height $z$ at radius $r$, $\Sigma_2(z,r)$
is the surface density {\em below} the point with height $z$ at radius $r$, and
$\zeta_0=1.3\cdot10^{-17}$~s$^{-1}$. Thus, we assume that cosmic ray particles
penetrate the disc only in the vertical direction from both its surfaces.
The X-ray ionisation rate $\zeta_\mathrm{X}$ is computed
according to Glassgold, Najita \&~Igea (\cite{zetaxa,zetaxb}) with parameters for
their high depletion case. The source of X-rays is assumed to be located at $z=12\,R_\odot$. 
We adopt the ionisation rate due to decay of radioisotopes to be 
$\zeta_\mathrm{R}=6.5\cdot10^{-19}$~s$^{-1}$.

The region under investigation spans a range of radii from 1 to 373~AU. We do not consider
regions closer to the star to avoid the necessity to account for physical
processes like dust destruction and collisional ionisation.
Even when the close vicinity of the star is excluded, the disc is still characterised
by a wide range of all relevant physical parameters: gas temperatures $T_\mathrm{g}$ are $10-600$~K,
densities are $10^{-20}-10^{-10}$~g~cm$^{-3}$, $G$ are $10^2-10^8$, $A_\mathrm{V}$ varies from 0 to more than 
100~mag and ionisation rates are $10^{-18}-10^{-10}$~s$^{-1}$.

We divide the disc into three regions, namely,
the midplane, the intermediate layer, and the surface layer.
This division is similar to that outlined by Aikawa \& Herbst (\cite{ah1999}). 
The midplane is the part of the disc near the symmetry plane, which is opaque
to both X-ray and UV photons and is ionised primarily by cosmic ray particles.
If one goes up from the midplane, the fractional ionisation first stays the
same as in the midplane, but then at some height it starts to grow in
response to increasing X-ray intensity. We take this turn-off
height to be the lower boundary of the intermediate layer (and the upper
boundary of the midplane region). In our model this height is approximately
given by $z_{\mathrm L}=0.05\,r^{1.42}$~AU. With increasing height,
the fractional ionisation grows first slowly, then much faster, in response
to the decreasing opacity of the medium. The point where transition between
slow and fast growth occurs is assumed to be the upper boundary of the
intermediate layer and the lower boundary for the third, surface layer.
This boundary is approximated by $z_{\mathrm U}=0.17\,r^{1.27}$~AU.
The fractional ionisation is important in dynamical calculations
only if it is small, as high ionisation means that the ideal MHD equations can be used.
Thus, we set the upper boundary of the surface layer to a position where the ionisation
degree exceeds $\sim10^{-4}$. This simple picture may not
be appropriate if diffusion processes or radial transport of the disc matter are taken 
into account (Ilgner et al.~\cite{IHMM03}).

\begin{figure}
\begin{center}
\includegraphics[width=0.45\textwidth,clip=]{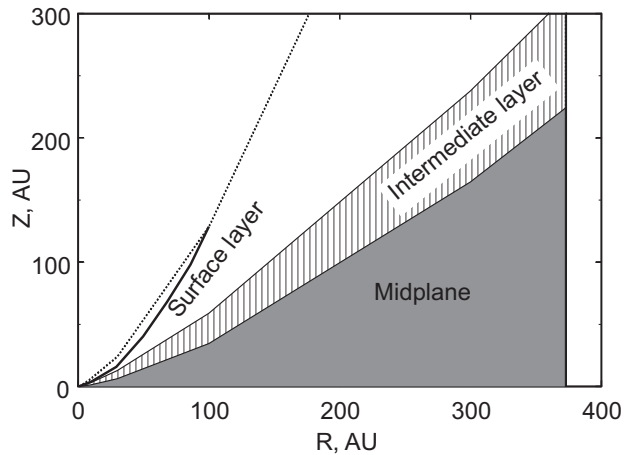}
\end{center}
\caption{Three layers of a disc with distinctively different sets
of chemical processes responsible for the value of the fractional
ionisation $x_\mathrm{e}$. The thick solid line is the upper
boundary of the surface layer where $x_\mathrm{e}=10^{-4}$, while dotted line depicts the disc upper limits.}
\label{dstruct}
\end{figure}

\begin{figure}
\begin{center}
\includegraphics[width=0.48\textwidth,clip=]{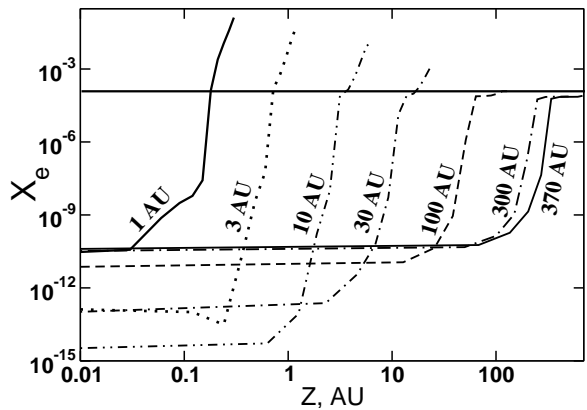}
\end{center}
\caption{Fractional ionisation as a function of height above the disc plane.
Lines are labelled with the corresponding radii.}
\label{idiagr}
\end{figure}

The disc structure is depicted in Figure~\ref{dstruct}. The upper boundary of the surface
layer is shown schematically (thick solid line), as at $R\ga100$~AU it extends beyond the disc limits
(dotted line) for the adopted cut-off fractional ionisation of $10^{-4}$. In Figure~\ref{idiagr}
we show the dependence of the fractional ionisation on the height above the disc plane computed
with the chemical model, described below, for several representative radii $R$.

\subsection{Chemical model}

The chemical model used in this paper is essentially the same as in Paper~I, but with a few corrections
that are described in this subsection.
We adopt the UMIST\,95 ratefile (Millar et al.~\cite{umist95}) for the gas-phase chemistry and the surface reaction
set compiled by Hasegawa, Herbst \& Leung~(\cite{hhl92}) and Hasegawa \&
Herbst~(\cite{hh93}). As we have multiple ionisation sources 
in the protoplanetary disc, the cosmic ray ionisation rate is replaced by the 
sum $\zeta_\mathrm{CR}+\zeta_\mathrm{X}+\zeta_\mathrm{R}$ in expressions for rates of
chemical reactions with cosmic ray particles (CRP) and cosmic ray induced photoprocesses.
Given the high intensity of the impinging UV radiation, photodesorption of surface
species is taken into account along with other desorption mechanisms.

\subsubsection{Grain charge}

The major difficulty in modelling the fractional ionisation in a dark and dense environment,
like a protoplanetary disc midplane, is that 
grains play a very important role in ion recombination and can be the dominant 
charged particles in some disc regions. The charge of grains and, consequently, their impact on the
local dynamics can sensitively depend on the grain size distribution, icy mantle composition 
etc. (e.g. Nishi, Nakano \& Umebayashi~\cite{Nishi}). Thus, the computation of the fractional 
ionisation in the disc midplane is related not only to the chemistry but to the grain physics as well. 
This means we should include at least a simplified treatment of grain charging processes in the model.

The following changes are made with respect to Paper~I. We consider neutral,
positively charged, and negatively charged grains 
separately. A grain attains a negative charge or loses a positive charge colliding with free electrons. 
The probability of an electron sticking to grain surfaces is assumed to decrease exponentially with dust
temperature $T_\mathrm{d}$, so that the sticking coefficient is around 0.5 at $T_\mathrm{d}\sim10-20$~K, 
close to the value obtained by Umebayashi \& Nakano (\cite{un1980}), and is essentially zero at higher
temperatures. Via dissociative recombination of a gas-phase ion, a grain loses one electron and becomes 
neutral or positively charged. Rates of collisions between electrons and positively charged grains are
multiplied by a factor, correcting for the long-distant Coulomb attraction, 
$C_{\rm ion}=1+e^2/ka_{\rm d}T_{\rm d}$, where $a_\mathrm{d}=0.1~\mu$m is the grain radius.
It is always assumed that $T_{\rm g}=T_{\rm d}=T$.

\subsubsection{Sticking probability}

In Paper~I the sticking probability $S$ was taken to be 0.3 at $T_{\rm g}=10$~K for
all neutral species except H, He, and H$_2$. However, at higher temperatures $S$
is likely to be smaller (Burke \& Hollenbach~\cite{BH1983}). To account for this
tendency, we multiply $S$ for neutrals by an additional factor equal to the fraction of
molecules of a given kind that have thermal energy lower than the desorption
energy $E_{\rm D}$ for this species (assuming a Maxwellian velocity distribution).

\subsubsection{Desorption processes}

The disc model, used in this paper, is characterised by a much wider range of physical
conditions than the model for a molecular cloud considered in Paper~I. This means, in particular, that
changes must be made to the adopted model of the species desorption from grain
surfaces. First, we now have the UV radiation in the model, so we add
photodesorption to the processes listed in Paper~I. The rate of photodesorption
is given by 
\begin{equation}
k_{\rm ph}=\left[G_{\mathrm{S}}\exp(-2A_\mathrm{V}^{\rm S})+G_{\mathrm{IS}}\exp(-2A_\mathrm{V}^{\rm IS})\right]
Y\pi a_{\rm d}^2,
\end{equation}
where $G_{\mathrm{S}}$ and $G_{\mathrm{IS}}$ are intensities of the stellar and interstellar UV 
radiation expressed in units of the mean interstellar radiation
field, $A_\mathrm{V}^{\rm S}$ and $A_\mathrm{V}^{\rm IS}$ are the corresponding 
visual extinctions in the direction to the star and in the vertical direction, 
and $Y$ is the photodesorption yield for which we adopted the expression
\begin{equation}
Y=0.0035+0.13\exp(-336/T_{\rm d}),
\end{equation}
derived by Walmsley, Pineau des For\^ets \& Flower~(\cite{wpf1999}) from the experimental data obtained by
Westley et al.~(\cite{westley}).

The cosmic ray desorption rate also needs to be revised. The expression
suggested by Hasegawa \& Herbst~(\cite{hh93}) and used in Paper~I is
based on the assumption that a cosmic ray particle (usually an iron
nucleus) deposits on average 0.4~MeV into a dust grain of the adopted
radius, impulsively heating it to a peak temperature $T_{\rm crp}$,
which is close to 70~K for cold $0.1\mu$m silicate dust. As in a disc
we expect much higher dust temperatures, data from L\'eger, Jura \&
Omont~(\cite{leger}) are used to take into account the dependence of
the cosmic ray heating on the initial dust temperature. The peak
temperature is approximated as
\begin{equation}
T_{\rm crp}=(4.36\cdot10^5+T_{\rm d}^3)^{1/3}.
\end{equation}
This expression predicts that a grain heats up to 76~K at $T_{\rm d}=10$~K and
gives almost no heating for $T_{\rm d}\ga100$~K.

\section{Initial conditions for chemistry}
\label{icon}

The problem of the initial conditions for chemical models of star-forming 
regions remains an open issue. The modelling of
pre-protostellar objects commonly starts with the purely atomic or
partly ionised initial composition, while in reality the initial abundances
even for a moderately dense medium must reflect its previous evolution which
ideally would have started from a diffuse intercloud gas. Even though
the atomic or ionic initial composition can be adequate in the prestellar or
pre-protostellar objects, the situation is different in protoplanetary discs,
where the chemical composition at the beginning of the modelling is obviously
far more advanced. One way to cope with this problem is to use ``sliding'' initial
conditions, when the initial abundances are set to be mostly atomic at the
outer radius of the disc. Then, a chemical model is run, and the final abundances
at a given radius are used as initial abundances at the next radius, closer to
the disc centre (Bauer et al.~\cite{bauer}; Willacy et al.~\cite{willacy}).
Another approach is to compute the input abundances with the model of a dark
cloud, out of which the disc has evolved (e.g. Aikawa \& Herbst~\cite{ah1999};
Aikawa et al.~\cite{warm}). Willacy et al.~(\cite{willacy}) compared both methods
and found that, with a few exceptions, they provide similar results, because in
a high-density environment
reasonable molecular abundances are reached within a fraction of a year even with
the atomic initial composition.

To simplify the reduction, we adopted the second approach. The choice of
a cloud model in Aikawa et al.~(\cite{warm}) and in previous papers of this
group is based on the analysis of the Ohio New Standard Model performed by
Terzieva \& Herbst~(\cite{th1998a}, \cite{th1998b}). These authors developed
a simple composite criterion that allows one to estimate
how good the chemical model is in reproducing the chemical composition of a given
object. Adopting a fixed density of $2\cdot10^4$~cm$^{-3}$ and a gas temperature of
10\,K, Terzieva \& Herbst~(\cite{th1998a}) found that with the pure gas-phase chemistry,
the chemical composition of the two typical molecular clouds, namely, TMC-1 and L134N,
is best reproduced after about $3\cdot10^5$~years of the evolution. Aiming to
reproduce the observed low abundances of water and molecular oxygen, Roberts \&
Herbst~(\cite{rh2002}) added surface chemistry to this model and found that the
``best consistency'' interval can be shifted toward a somewhat later time, like a few
times $10^6$~years.

\begin{figure}
\begin{center}
\includegraphics[width=0.5\textwidth,clip]{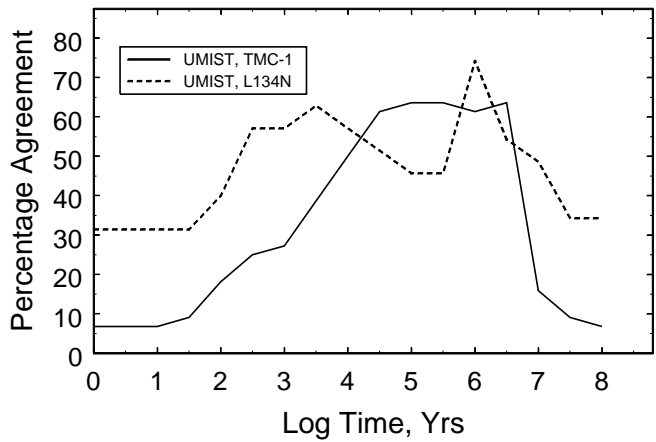}
\end{center}
\caption{Percentage agreement between calculated and observed gas-phase abundances.}
\label{agree}
\end{figure}

As we use the UMIST\,95 ratefile, we perform a similar analysis for this
database. The observed abundances for TMC-1 and L134N are taken from
Ohishi, Irvine \& Kaifu~(\cite{ohishi}) and Langer et al.~(\cite{h2c6}). Following Terzieva \& 
Herbst~(\cite{th1998b}), we assume that the abundance of a given molecule is consistent with 
observational data if it lies within an order of magnitude of the observed abundance. When
the observed abundance is an upper limit, we assume that the computed
abundance is in agreement with observations if it is less than or equal
to 10 times this upper limit. With the same density, temperature and
the surface chemistry included, we find that for our chemical model the
agreement is best at $t\sim10^6$~years (Fig.~\ref{agree}). The ``low-metal''
set of abundances from Paper~I is adopted for this computation.

We now use the set of gas-phase
and surface abundances, obtained with the above model for $t=10^6$~years as the initial conditions.
The brief outline of this set is the following (abundances relative to the number of
hydrogen nuclei are given in parentheses): The primary carbon-bearing compound
is surface formaldehyde ($3.7\cdot10^{-5}$) which also accounts for a significant
fraction of the overall oxygen content, being second only to surface water
($7.8\cdot10^{-5}$). Gas-phase oxygen and carbon are locked in CO molecules
($2.8\cdot10^{-5}$). A fraction of oxygen is also present in atomic ($1.1\cdot10^{-5}$)
and molecular ($8.7\cdot10^{-6}$) form. Gas-phase nitrogen is contained mainly
in N$_2$ molecules ($9.7\cdot10^{-6}$), while in icy mantles the main N-bearing species
are HCN and HNC ($\sim10^{-6}$ each) with a slightly lower amount of CH$_2$NH, NH$_3$,
and NO. Sulphur is distributed almost equally between gas and dust phases. In the
gas-phase it is locked in the CS molecules ($2.9\cdot10^{-8}$). Surface sulphur is bound
in H$_2$S ($2.2\cdot10^{-8}$) and H$_2$CS ($1.5\cdot10^{-8}$). Even though the
density is not very high, metals are already significantly depleted. Surface
abundances of Mg, Fe, and Na constitute about 1/3 of their total abundances.
The dominant ions are HCO$^+$ ($3.8\cdot10^{-9}$), H$_3^+$ ($3.5\cdot10^{-9}$),
Mg$^+$ ($1.9\cdot10^{-9}$), C$^+$ ($1.3\cdot10^{-9}$), and Fe$^+$ ($1.2\cdot10^{-9}$).

It should be kept in mind that some of our reduced networks computed for this
initial composition may not be appropriate for purely atomic initial abundances
of for abundance sets with higher metal content.

\begin{table*}
\caption{Dominant ions in the midplane, intermediate layer and surface layer at $t=10^6$~years.}
\label{domions}
\begin{tabular}{llllllll}
\hline
R, AU    &  1      &  3  &  10  &  30  &  100  &  300  & 373 \\
\hline
         &         & &         &         &            &            & \\
Midplane & Na$^+$  & HCNH+      & HCO$^+$ & HCO$^+$ & N$_2$H$^+$ & H$_3^+$ & H$_3^+$ \\
         &         & &         &         &            &            & \\
Intermediate   & Mg$^+$  & HCO$^+$ & HCO$^+$ & HCO$^+$ & HCO$^+$    & HCO$^+$ & HCO$^+$ \\
layer   &         &  S$^+$   &         &         &            &            & \\
         &         & H$_3^+$    &         &         &            &            & \\
         &         & and others&         &         &            &            & \\
         &         & &         &         &            &            & \\
Surface  & C$^+$   & C$^+$      & C$^+$   & C$^+$   & C$^+$      &  C$^+$     & C$^+$ \\
layer & H$^+$   & H$^+$      &         &         &            &            & \\
\hline
\end{tabular}
\end{table*}

\section{Chemistry of the ionisation degree in a protoplanetary disc}
\label{res}

In each of the selected disc points the chemical model described in Section~\ref{mod}
is run up to $t=10^6$~years. Dominant ions in various disc domains are
listed in Table~\ref{domions}. Reduction methods are then used to find
those processes that determine $x_{\rm e}$. All the reduced networks
described in this section are freely available from the authors.
Also, we show some representative networks for the midplane and the
surface layer.

\subsection{Ionisation in the midplane}

The midplane of the disc is characterised by high density, high optical depth
and a relatively low ionisation rate. Neither the UV radiation
from a central star nor the interstellar UV radiation is able to penetrate into
this region. The ionisation rate is close to $10^{-17}$~s$^{-1}$. At the point
closest to the star, $\zeta$ is dominated by X-ray ionisation; further away
from the centre the attenuated cosmic rays are the only ionising factor.
The temperature drops from $\sim600$~K at $R=1$~AU to about 8~K at the disc edge.
In the same distance interval, the gas density goes down from $10^{-10}$~g~cm$^{-3}$
to less than $10^{-14}$~g~cm$^{-3}$. Detailed physical conditions for the midplane
are given in Table~\ref{midcond}.

\begin{table}
\caption{Physical conditions in the midplane.}
\label{midcond}
\begin{tabular}{lrcrcc}
\hline
No. & $R$, AU & $\rho$, g cm$^{-3}$ & $T$, K & $\zeta_{\rm CR}$,~s$^{-1}$ & $\zeta_{\rm X}$,~s$^{-1}$\\
\hline
M1 & 1.  &  2.1($-$10) & 614. &  8.8($-$19) & 7.5($-$18) \\
M2 & 3.  &  4.7($-$11) & 193. &  2.3($-$18) & 4.0($-$23) \\
M3 & 10. &  8.5($-$12) &  52. &  5.0($-$18) & 1.2($-$25) \\
M4 & 30. &  7.3($-$13) &  29. &  9.4($-$18) & 5.4($-$27) \\
M5 & 100.&  4.8($-$14) &  16. &  1.2($-$17) & 3.8($-$28) \\
M6 & 300.&  4.8($-$15) &   9. &  1.3($-$17) & 4.0($-$29) \\
M7 & 370.&  3.3($-$15) &   8. &  1.3($-$17) & 2.6($-$29) \\
\hline
\end{tabular}
\end{table}

\subsubsection{Dark, hot chemistry}

\begin{table}
\caption{Reduced network for dark, hot chemistry in the midplane.}
\begin{tabular}{lcl}
\hline
${\rm NO}^+ + {\rm Mg}$&$ \rightarrow $&${\rm Mg}^+ +   {\rm NO} $   \\
${\rm Mg}^+ +   e^- $&$ \rightarrow $&${\rm Mg} + h\nu$ \\
${\rm NO} +  h\nu_{\rm CR}  $&$ \rightarrow $&$ {\rm NO}^+ + e^-$\\
\hline
\end{tabular}
\label{m1}
\end{table}

The chemistry of the point M1 (see Table~\ref{midcond}) is extremely simple. In essence, it involves
only five species, which are neutral and ionised magnesium, neutral and
ionised NO, and electrons. The reduced network for this point is
shown in Table~\ref{m1}. At the beginning of the computation the electron abundance
decreases from the initial value of $\sim 10^{-9}$ to the new equilibrium value of about $10^{-11}$, 
determined by the electron exchange between Mg, Mg$^+$, NO, and NO$^+$, and then remains
nearly constant. With the full network, the dominant ion is Na$^+$, but it serves mainly as a 
passive sink for positive charges. This is why it is taken out of the reduced network. 
It suffices to include ionised magnesium, which is one of the most abundant ions in the initial
abundance set.

Note the apparent similarity of this network to the one
that determines $x_{\rm e}$ in dense interstellar clouds (Oppenheimer \& Dalgarno
\cite{OD74}) in the sense that it consists of a representative metal, a
representative molecular ion and the charge transfer reaction between them.
The explanation is that, even though we include gas-grain
interaction in our model, at such high temperature the chemistry
mostly proceeds in the gas-phase. Thus, the ionisation degree is determined
by gas-phase reactions. However, due to much higher density these reactions
are different from those described in Oppenheimer \& Dalgarno (\cite{OD74}).
The major difference is that the reaction of H$_2$ ionisation by cosmic rays
is not important in the considered point. A modest electron supply is provided
by NO ionisation by cosmic ray induced photons. The
importance of the NO$^+$ ion is determined by its chemical inertness. Due to this
inertness, most of the time its abundance exceeds that of other ions, involved in charge 
transfer reactions with magnesium, by more than an order of magnitude. With this $5\times3$ 
network, the error in the fractional ionisation does not exceed 40\% during the entire computation 
time. By adding a few more species to the reduced network, it is possible to
decrease this uncertainty to 20\%. These species account for the rapid
destruction of the two other initially abundant ions, i.e. HCO$^+$
and H$_3^+$.

The chemistry of the point M2 is also extremely simple, but in a totally
different way. At this somewhat lower temperature, $T\sim 200$~K, magnesium is irreversibly 
depleted onto dust grains. While at earliest times ($t<3$~years) the evolution of the electron
abundance is governed by recombination of Mg$^+$, later the
fractional ionisation is equilibrated and determined by a simple network involving H, H$_2$,
H$_2^+$, and H$_3^+$. Included reactions are H$_2$ ionisation by cosmic rays,
H$_3^+$ formation, H$_3^+$ and Mg$^+$ recombination, Mg accretion onto dust
grains, and grain (re)charging processes. The error in the fractional ionisation does not 
exceed 10\% during almost the entire computation time, reaching 40\% only at the very beginning of the
computation due to neglected Na$^+$ and Fe$^+$ ions. Note that the equilibrated gas-phase electron 
abundance $x_\mathrm{e}\sim10^{-13}$ at this point is several times lower than the abundance of 
charged grains. Thus the net charge density at these conditions is determined not by the 
chemical kinetics, but mainly by grain charging processes.

Dominant ions in a model of a disc interior which is similar to the
one used here are determined by Markwick et al.~(\cite{markwick}). In the second
part of their Table~1 they present ions that define the $x_{\rm e}$ value for
various heights and at $R\le10$~AU, when X-rays are taken into account.
The difference to our results is obvious. In our model, the dominant ions throughout
the inner part of the midplane are Na$^+$, HCNH$^+$, and HCO$^+$, while in the model
by Markwick et al.~(\cite{markwick}) everywhere in the midplane the fractional
ionisation is determined by the rather complex CH$_3$CO$^+$ ion. We attribute this
discrepancy to the different treatment of the surface dissociative recombination of
ions as well as to different treatment of grain charges. In the Markwick et al. model
all grains are assumed to participate in recombination reactions, with products of
recombination sticking
to grain surfaces, while in our model these products are returned to the gas immediately.
So, in their model, metal ions probably recombine with grains and
stick to dust surfaces, which implies that they are not easily returned to
the gas-phase. In our model, in the point
closest to the star, all dust grains are positively charged, so they do not contribute to
the recombination rate at all.
Also, when a metal ion happens to encounter a rare negative or neutral grain, it loses
the charge but remains in the gas-phase. Higher above the midplane and somewhat further
from the star, where details of the gas-grain interaction are less
important, the dominant ions in both models are almost the same, i.e., HCNH$^+$ and
HCO$^+$.

\subsubsection{Dark, cold chemistry}

The ionisation at the points M3 and M4 displays similar behaviour. In a cold and dense
environment, metals are depleted onto
dust grains almost immediately, and the gas-phase fractional ionisation drops to a very low equilibrium
value -- about $3\cdot10^{-15}$ in M3 and about $8\cdot10^{-14}$ in M4. These values are
determined by three almost equally important cycles, ionisation and recombination
of helium, formation and destruction of HCO$^+$ and formation and destruction of H$^+_3$.
The reduced network, consisting of these species and processes, reproduces the fractional
ionisation with a few per cent accuracy in all but the earliest times. The electron abundance
at $t<0.1$~yrs in the absence of metals is determined by a more complicated set of processes,
with the fractional ionisation being reproduced within 50\% uncertainty by a network of about
40~species. Again, the total charge density at these points is determined by charged grains,
not by gas-phase ions and electrons.

The dominant ions at points M5, M6, and M7 are N$_2$H$^+$, HCO$^+$ and H$_3^+$.
The reduced  network for the entire outer part of the disc consists of
about 25~species and of a comparable number of reactions, which govern abundances 
of these ions. At earlier times Mg$^+$ and Fe$^+$ are important as electron suppliers,
so in addition to N$_2$H$^+$ and HCO$^+$ chemistry, the reduced 
network includes also neutral and ionised metals. Among the included reactions
are metal sticking to grains, 
cosmic ray ionisation of molecular hydrogen, H$_3^+$ formation, charge transfer between metals 
and H$_3^+$, and dissociative recombination of ions with grains and electrons.
The network responsible for $x_{\rm e}$ at the outer disc edge is presented
in Table~\ref{m7}. It also includes reactions of adsorption and desorption of neutral
components that are not shown for brevity.

\begin{table}
\caption{Reduced network for dark, cold chemistry in the midplane.}
\begin{tabular}{lcl}
\hline
${\rm H}_2 + {\rm CRP}  $&$\rightarrow$  &$  {\rm H}_2^+   +  e^-$\\
${\rm H}_2^+  +   {\rm H}_2     $&$\rightarrow$  &$ {\rm H}_3^+ + {\rm H}$\\
${\rm H}_3^+   +  {\rm N}_2     $&$\rightarrow$  &$ {\rm N}_2{\rm H}^+  +  {\rm H}_2$ \\
${\rm H}_3^+ +   e^-  $&$\rightarrow$  &$ {\rm H}_2    +  {\rm H}$\\
${\rm HCO}^+  +  e^-  $&$\rightarrow$  &${\rm CO}  +  {\rm H}$\\
${\rm N}_2{\rm H}^+  +  e^- $&$\rightarrow$  &${\rm N}_2   +   {\rm H}$\\
${\rm Fe}^+ +    g^-  $&$\rightarrow$  &${\rm Fe}     $\\
${\rm Mg}^+   +  g^-  $&$\rightarrow$  &${\rm Mg}  $\\
${\rm H}_3^+ +   g^-  $&$\rightarrow$  &$ {\rm H}_2    +  {\rm H}$\\
${\rm HCO}^+  +  g^-  $&$\rightarrow$  &${\rm CO}  +  {\rm H}$\\
${\rm N}_2{\rm H}^+  +  g^- $&$\rightarrow$  &${\rm N}_2   +   {\rm H}$\\
\hline
\end{tabular}
\label{m7}
\end{table}

\subsection{Ionisation in the intermediate layer}

The intermediate layer differs from the midplane by a much higher X-ray
ionisation rate. Ranges of physical conditions for the intermediate layer
are summarised in Table~\ref{intermed}. They imply that many molecules
are abundant in the gas phase in this disc domain due to shielding of the UV
radiation and effective desorption by X-rays. From the chemical point of 
view, the characteristic feature of the intermediate layer is that a
chemo-ionisation equilibrium is sometimes not reached during the
entire 1~Myr time span considered.

\subsubsection{Warm, X-ray driven chemistry}

Similar to the midplane, the part of the intermediate layer closest to the
star ($R<3$~AU) is characterised by temperatures of a few hundred~K.
As the grain icy mantles are nearly absent, the chemistry in
this region is driven by gas-phase reactions. Dominant ions are
metals, HCO$^+$ and HCNH$^+$. Apart from these components, the reduced
network contains CO and neutral metals as well as neutral and ionised N and neutral O along 
with their hydrides  [N,O]H$_n$ and [N,O]H$_n^+$, involved in synthesis and destruction of the
hydrogen-saturated molecules NH$_3$ and H$_2$O. Also included are N$_2$, He, He$^+$, which are 
needed closer to the bottom of the layer (where $T$ is high and $\zeta$ is relatively
low), and HCN, HNC, HCN$^+$, O$_2$, and O$_2^+$ which are important further away from the
midplane (where $T$ is low and $\zeta$ is very high). The fractional ionisation
computed with this reduced network differs from the one computed with the full
network by less than 40\% during the entire time span.

\begin{table}[t]
\caption{Physical conditions in the intermediate layer.}
\label{intermed}
\begin{tabular}{llll}
\hline
$R$, AU & $\rho$, g cm$^{-3}$ & $T$, K& $\zeta$,~s$^{-1}$ \\
\hline
1  & $4(-12)-2(-10)$&  $140-520$ & $3(-13)-4(-10)$ \\
3  & $4(-15)-6(-12)$&  $80-120$  & $2(-15)-2(-12)$ \\
10 & $1(-15)-4(-14)$&  $50-90$   & $8(-16)-2(-13)$ \\
30 & $6(-17)-1(-14)$&  $30-80$   & $4(-17)-6(-15)$ \\
100& $4(-18)-1(-16)$&  $30-60$   & $7(-17)-1(-15)$ \\
300& $2(-18)-2(-17)$&  $20-40$   & $1(-17)-7(-17)$ \\
370& $1(-18)-1(-17)$&  $20-40$   & $1(-17)-4(-17)$ \\
\hline
\end{tabular}
\end{table}

Again comparing our data for the intermediate layer to those by Markwick et al.
(\cite{markwick}) we find that at $3<R<10$~AU our results generally agree,
especially given the fact that the sampling of the inner part of
the disc is much more detailed in the Markwick et al. model. Again, in the innermost
point we find a dominant metal ion in our model compared to molecular ions
HCNH$^+$ and H$_4$C$_2$N$^+$ in theirs. We believe that the origin of this
difference is the same as described above.

\subsubsection{Cold, X-ray driven chemistry}

The ionisation degree further out from the star (at $R>30$~AU) is
established by the evolutionary sequence which consists of the following stages.
Initially, the more abundant HCO$^+$ and H$_3^+$ species are destroyed almost
immediately and attain abundances of the order of $10^{-11}$ at $t\sim10^{-3}$~years.
Metal ions are neutralised somewhat more slowly through recombination with negatively charged grains and electrons
and stick to dust surfaces. The overall electron concentration $x_\mathrm{e}$ decreases
from $\sim 3\cdot10^{-9}$~cm$^{-3}$ at $t=0$ to $10^{-9}$~cm$^{-3}$ at $t \sim 0.3$~yrs.

Then, few equilibrium states are reached with a sequence of molecular ions,
mainly, NH$_4^+$, H$_3$CO$^+$, and HCO$^+$. Equilibrium abundance of
the next ions in this sequence is higher and is reached later, thus, the time dependence of
the fractional ionisation has a step-like appearance (after the initial decrease).
The ``second birth'' of HCO$^+$ is caused by H$_2$CO coming from icy mantles.
Formaldehyde reacts with H$_3^+$ to form H$_3$CO$^+$, which is the dominant ion
at $0.1<t<30$~yrs. One of the recombination channels for H$_3$CO$^+$ maintains
its equilibrium, restoring H$_2$CO. The other two channels lead directly and 
indirectly to CO production and
eventually increase the abundance of this molecule up to the point where
a new equilibrium state is reached for HCO$^+$. This ion eventually dominates the fractional
ionisation in the outer part of the intermediate layer.
Direct formation of HCO$^+$ in the reaction of H$_3^+$ with CO is damped initially in
favour of the H$_3^+$ reaction with H$_2$CO both because of the lower initial CO abundance and
because the latter reaction has a higher rate coefficient $\alpha = 6.3\cdot10^{-9}$~cm$^3$\,s$^{-1}$ compared 
with $\alpha = 1.7\cdot10^{-9}$ for H$_3^+$ + CO reaction.

To reproduce the step-like $x_\mathrm{e}$ behaviour, one needs a reduced network
that includes a few tens of species involved in cycles of synthesis and destruction
of the above ions. This network reproduces the nature of the equilibrium stages and their
length with an uncertainty that does not exceed 20\%.

\begin{figure}
\begin{center}
\includegraphics[width=0.45\textwidth,clip]{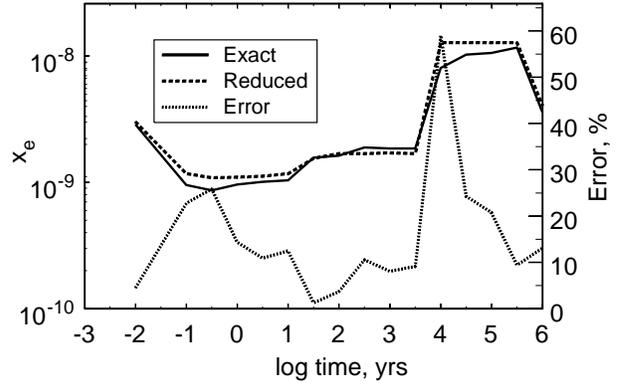}
\end{center}
\caption{Evolution of the fractional ionisation in the intermediate layer at
$R=3$~AU. The exact solution is shown by the solid line. The dashed line corresponds to the
reduced network consisting of 112 species and 195 reactions.
The dotted line (right Y-axis) shows the relative uncertainty in the
fractional ionisation computed with the reduced network.}
\label{fmed24}
\end{figure}

This simple picture is not appropriate closer to the star ($3<R<30$~AU),
where the density is high enough and the temperature is low enough to allow effective
gas-dust interaction. At the same time, a high ionisation rate leads to an increased
abundance of helium ions, which significantly alters the late-time evolution of
the fractional ionisation. An example of this evolution is shown in Figure~\ref{fmed24}.
It corresponds to the case with $n_\mathrm{H}=8.4\cdot10^{10}$~cm$^{-3}$, $T=74$~K,
$\zeta=2\cdot10^{-13}$. The ``exact'' solution is shown with the solid line.
The initial decrease of the electron abundance is followed by
a few steps corresponding to different equilibrium states. The last one, that of HCO$^+$,
is reached after approximately 300~yrs of evolution. If there was no relatively
abundant ionised helium, the fractional ionisation would stay constant from that moment. It is this ion,
coupled to high density and low temperature, that causes a new development.
At $t\sim3\cdot10^4$~yrs the ionisation degree increases by about an order
of magnitude up to $\sim 10^{-8}$~cm$^{-3}$, and another equilibrium
state is reached, with S$^+$ being a dominant ion and a dominant
sulphur-bearing species.

In the initial abundance set sulphur is mostly bound in CS and H$_2$S molecules,
which are destroyed by He$^+$, producing either S$^+$ or S (as well as C, C$^+$, and H$_2$).
Passing through a few reactions, the most important of which is the reaction with CH$_4$, 
sulphur is reunited with carbon in a CS molecule. This process is well
equilibrated, when the abundance of methane is high, and leads to a negligible
S$^+$ abundance at $t<10^4$~yrs. However, with time, some carbon atoms 
(both free and incorporated into
CH$_4$) are taken out of
this cycle due to formation of long carbon chains in reactions of C$^+$ with
methane and other hydrocarbons. After 3000~yrs the cycle is broken
entirely. As the abundance of methane and other light carbon-bearing species
goes down, that of ionised sulphur grows, causing almost an order
of magnitude increase in the fractional ionisation.

\begin{figure}
\begin{center}
\includegraphics[width=0.45\textwidth,clip]{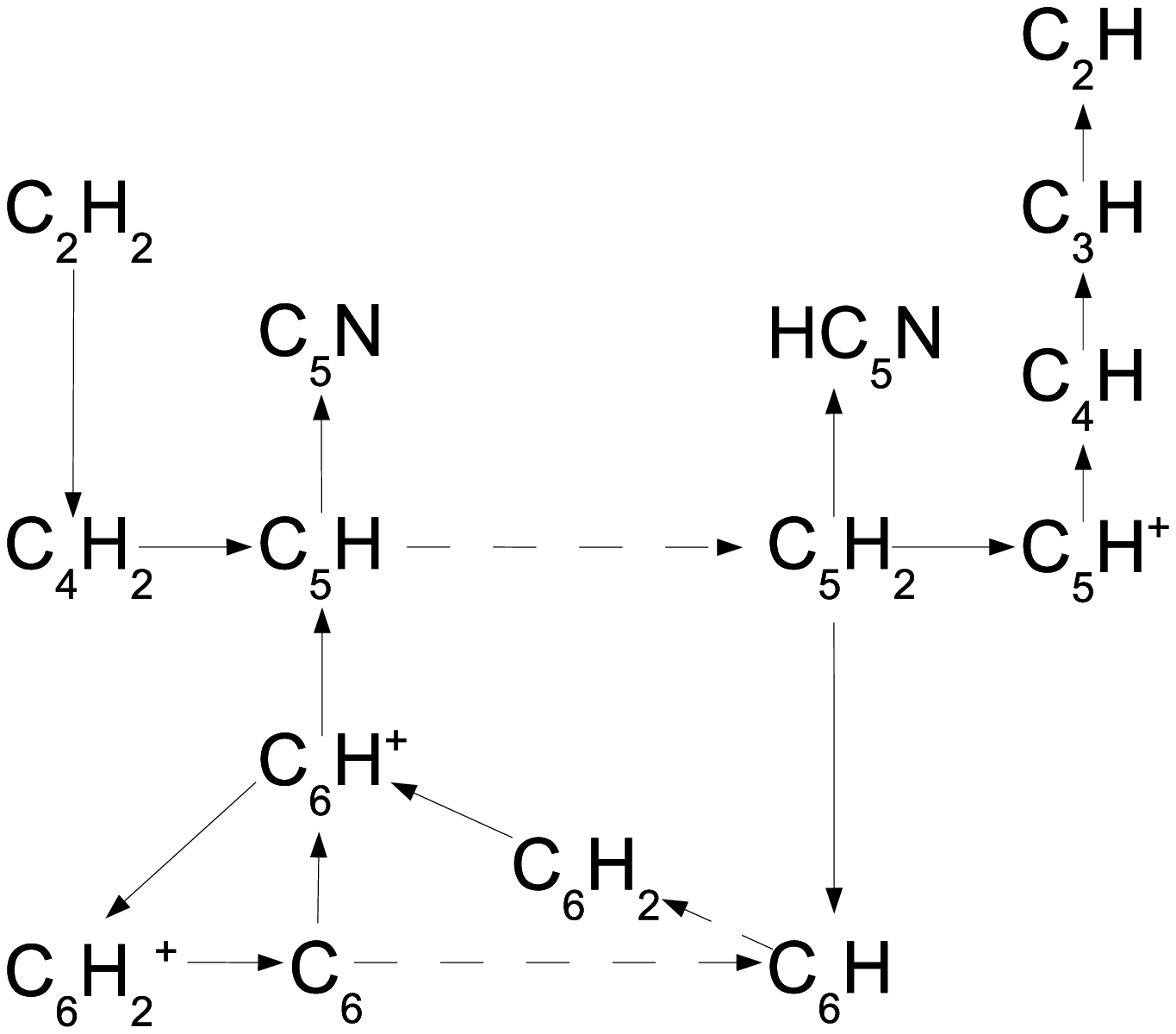}
\end{center}
\caption{Main routes governing synthesis and destruction of long carbon chains. Surface reactions
are indicated by dashed lines.}
\label{c6}
\end{figure}

Further evolution of $x_\mathrm{e}$ is implicitly controlled
by gas-phase and surface reactions involving carbon chains. The reduction
methods allow one to find which
of them are most important, as shown in Figure~\ref{c6}. The main route for synthesis
and destruction of long carbon chains in the depicted scheme starts with an acetylene
molecule containing only two
carbon atoms, which is rapidly converted to C$_5$H$_2$ molecules. This molecule
then either transforms to even longer chains, or proceeds further down the cycle to
C$_5$H$^+$ which recombines to C$_4$H. This species starts a chain of very
slow reactions with atomic oxygen that degrades it to C$_3$H, C$_2$H, and CH. Eventually,
destruction of carbon chains leads to the new growth of methane abundance and to the slow
decrease of the fractional ionisation. Ionised sulphur is consumed in the reaction
with CH$_4$ and eventually is converted back to CS. The abundance of H$_3^+$ determines the 
ionisation degree at the end of the computation.

Conversion of C$_5$H$_2$ molecules into longer chains (lower part of Figure~\ref{c6})
slows down the methane abundance
restoration. The reduced network must include almost all species with six carbon atoms to
reproduce properly the time scale of this process. Three key
reactions of this carbon chain cycle proceed on dust grain surfaces. These are
reactions of hydrogen addition to C$_5$H, C$_6$, and C$_6$H (shown with
dashed lines). Without these reactions the sequence of carbon chain synthesis would be 
terminated by the C$_5$N molecule that lacks effective destruction pathways in the given 
conditions. Exhaustion of atomic and ionised carbon
would keep the ionised sulphur at a high level, reached at $t\sim3000$~years.
Thus, we must conclude that at least within the limits of the adopted
chemical model some surface reactions are of crucial significance.
As far as we know, this is the first indication that the surface chemistry may
have an ``order-of-magnitude'' importance for the evolution of fractional
ionisation in interstellar matter.

\subsection{Ionisation in the surface layer}

The surface layer encompasses those disc regions where the degree of ionisation
reaches a value of $\sim 10^{-4}$ at the end of the evolutionary time. The
corresponding physical conditions imply low densities 
($\rho \sim 10^{-19}- 10^{-13}$~g cm$^{-3}$, moderate temperatures
($T < 250$~K), a high X-ray ionisation rate and low obscuration of the
UV radiation. Conditions in the surface layer are given in
Table~\ref{surface}. Note that the temperature of the disc atmosphere is 
nearly constant in the vertical direction at any distance from the star.
This rather extreme environment presumes a chemical evolution typical
of photon-dominated regions (PDR). The chemistry is dominated by gas-phase
processes, while the role of gas-grain interactions is negligible, apart from
the desorption of initially abundant surface species. As our definition of the
surface layer is based on the fractional ionisation value,
the layer can be further divided into two chemically distinct
zones. Closer to the star ($r\la10$~AU) $x_\mathrm{e}$ is controlled by the 
X-ray ionisation, while in the outer regions ($r\ga10$~AU) ionisation is mainly 
provided by the UV photons.

\subsubsection{X-ray dominated chemistry}

\begin{table*}
\caption{Physical conditions in the surface layer.}
\label{surface}
\begin{tabular}{llllllll}
\hline
$R$, AU & $\rho$, g cm$^{-3}$ & $T$, K& $G_0$& $A_{\rm V}^{\rm S}$,~mag & $A_{\rm V}^{\rm IS}$,~mag  & $\zeta$,~s$^{-1}$ \\
\hline
1    & $2(-14)-1(-13)$ & $\sim180$ & $1(8)$      & $>10$      & $\sim1$ & $2(-9)$ \\
3    & $2(-16)-1(-15)$ & $\sim230$ & $7(7)$      & $\sim1.0$  & $<0.5$  & $8(-12)$ \\
10   & $1(-17)-8(-17)$ & $140-160$ & $9(5)$      & $<1$       & $<0.5$  & $6(-13)$ \\
30   & $3(-18)-1(-17)$ & $80-100$ &  $9(4)-1(3)$ & $<1$       & $<0.5$  & $5(-14)$ \\
100  & $4(-19)-4(-18)$ & $60-70$  &  $5(3)-7(3)$ & $<1$       & $<0.5$  & $1(-15)-4(-15)$ \\
300  & $3(-19)-2(-18)$ & $35-40$  &  $5(2)-7(2)$ & $<1$       & $<0.5$  & $7(-17)-2(-16)$ \\
370  & $1(-19)-1(-18)$ & $30-40$  &  $2(2)-4(2)$ & $<1$       & $<0.5$  & $5(-17)-1(-16)$ \\
\hline
\end{tabular}
\end{table*}

\begin{table}
\caption{Reduced network for X-ray dominated chemistry in the surface layer
(superscript ``d'' is used to denote surface species).}
\begin{tabular}{lcl}
\hline
${\rm H}_2{\rm CO}^{\rm d}$&$\rightarrow$  &${\rm H}_2{\rm CO}$\\
${\rm H}_2{\rm O}^{\rm d}$&$\rightarrow$  &${\rm H}_2{\rm O}$\\
${\rm C}   +    h\nu_{\rm CR}$&$\rightarrow$  &${\rm C}^+    +  e^-$\\
${\rm OH}  +    h\nu_{\rm CR}$&$\rightarrow$  &${\rm O} +      {\rm H}           $\\
${\rm CO}  +    h\nu_{\rm CR}$&$\rightarrow$  &${\rm C}  +     {\rm O}           $\\
${\rm H}_2{\rm CO} +   h\nu_{\rm CR}$&$\rightarrow$  &${\rm CO} +     {\rm H}_2          $\\
${\rm H}   +    $\,CRP,X-ray&$\rightarrow$  &${\rm H}^+   +   e^-      $\\
${\rm H}_2  +$\,CRP,X-ray&$\rightarrow$  &${\rm H}_2^+    + e^-      $\\
${\rm H}_2^+ +    {\rm H}_2    $&$\rightarrow$  &${\rm H}_3^+   +  {\rm H}           $\\
${\rm H}^+  +    {\rm O}     $&$\rightarrow$  &${\rm O}^+     + {\rm H}           $\\
${\rm H}^+  +    {\rm OH}    $&$\rightarrow$  &${\rm OH}^+    + {\rm H}           $\\
${\rm H}_3^+ +    {\rm O}    $&$\rightarrow$  &${\rm OH}^+    + {\rm H}_2          $\\
${\rm H}_3^+ +    {\rm H}_2{\rm O}   $&$\rightarrow$  &${\rm H}_3{\rm O}^+  +  {\rm H}_2          $\\
${\rm H}_3^+ +    {\rm CO}   $&$\rightarrow$  &${\rm HCO}^+   + {\rm H}_2          $\\
${\rm O}^+  +    {\rm H} $&$\rightarrow$  &${\rm H}^+    +  {\rm O}           $\\
${\rm O}^+  +    {\rm H}_2  $&$\rightarrow$  &${\rm OH}^+   +  {\rm H}           $\\
${\rm OH}^+ +    {\rm H}_2  $&$\rightarrow$  &${\rm H}_2{\rm O}^+  +  {\rm H}           $\\
${\rm H}_2{\rm O}^++    {\rm H}_2 $&$\rightarrow$  &${\rm H}_3{\rm O}^+  +  {\rm H}           $\\
${\rm C}^+  +    {\rm H}_2{\rm O}   $&$\rightarrow$  &${\rm HCO}^+ +   {\rm H}           $\\
${\rm C}^+  +    {\rm O}_2    $&$\rightarrow$  &${\rm O}^+   +   {\rm CO}          $\\
${\rm O}   +    {\rm OH}   $&$\rightarrow$  &${\rm O}_2  +    {\rm H}           $\\
${\rm H}^+  +    e^-$&$\rightarrow$  &${\rm H}      + h\nu    $\\
${\rm H}_3^+ +    e^-$&$\rightarrow$  &${\rm H}_2    +  {\rm H     }      $\\
${\rm C}^+  +    e^-$&$\rightarrow$  &${\rm C}    +   h\nu      $\\
${\rm H}_2{\rm O}^++    e^-$&$\rightarrow$  &${\rm O}  +     {\rm H}_2          $\\
${\rm H}_3{\rm O}^++    e^-$&$\rightarrow$  &${\rm OH}  +    {\rm H}  +   {\rm H}  $\\
${\rm HCO}^++    e^-$&$\rightarrow$  &${\rm CO}   +   {\rm H}           $\\
\hline
\label{m1_6}
\end{tabular}
\end{table}

\begin{figure}
\begin{center}
\includegraphics[width=0.5\textwidth,clip]{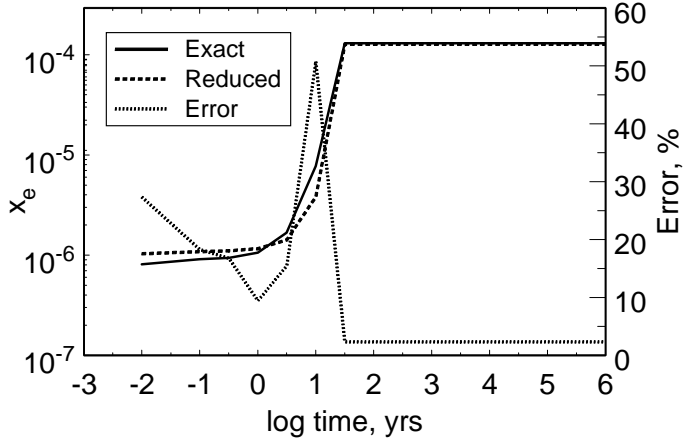}
\end{center}
\caption{Evolution of the fractional ionisation in the surface layer at
$R=1$~AU. The exact solution is shown with the solid line. Dashed line corresponds to the
reduced network consisting of 21 species and 27 reactions.
Dotted line (right Y-axis) shows the relative uncertainty in the
fractional ionisation computed with the reduced network.}
\label{fmed16}
\end{figure}

The evolution of the ionisation degree in the hot ($T\sim 200$~K) part of the surface
layer closer to the star is determined by X-ray ionisation of atomic hydrogen. The
final equilibrium ionisation degree is well reproduced by the network, consisting of
hydrogen-bearing species (H, H$^+$, H$_2^+$, H$_3^+$) only. This equilibrium
value is reached after 100~years of evolution. Up to this time $x_\mathrm{e}$ is
determined by a more complicated set of chemical processes, which includes H$_3^+$
formation, an H$_3$O$^+$ formation and destruction cycle as well as reactions involving
CO and HCO$^+$. The primary carbon and oxygen carrier, formaldehyde, adds to this
complexity by being desorbed from grain surfaces and destroyed by cosmic ray and X-ray
induced photons. The cycles that involve these molecules are not equilibrated. Most
reactions with polyatomic species tend to produce atomic hydrogen, which is 
gradually ionised by intense X-ray radiation. Eventually, all complex molecules
are destroyed, and at the end of the computation, the gas mostly consists of atoms and atomic
ions. The entire reduced network comprises 21 species and 27 reactions
and is presented in Table~\ref{m1_6}.
The uncertainty does not exceed 60\% during the entire computational time
(Fig.~\ref{fmed16}). Without this simple, but important
ion-molecule chemistry, the uncertainty at earlier times would be much larger
(exceeding an order of magnitude).

\subsubsection{UV-dominated chemistry}

The evolution of the ionisation fraction in the distant regions of the surface layer
($r\ga10$~AU) is similar but more complicated from the chemical point of view
than in the case of the X-ray dominated chemistry. As the intensity of stellar
X-rays decreases, UV photons become the main ionising source
(see Table~\ref{surface}). The typical gas temperature in this part of the disc
decreases below $100$~K. The entire ionisation evolution of the region is determined
by formaldehyde desorption and destruction. Icy mantles illuminated by UV radiation
evaporate, thus delivering H$_2$CO into the gas phase. This molecule is either
destroyed directly by photons (H$_2\mathrm{CO}\rightarrow\mathrm{CO,HCO}^+$) or
ionised and dissociated (H$_2\mathrm{CO}\rightarrow\mathrm{H}_2\mathrm{CO}^+
\rightarrow\mathrm{CO}$). HCO$^+$ dissociatively recombines to CO as well, and this molecule
is dissociated by photons producing C and O. The destruction processes proceed
very fast, so after only a fraction of a year the gas composition is mostly atomic,
with C$^+$ being the dominant ion. When all polyatomic species are destroyed,
the fractional ionisation is regulated by an equilibrium ionisation-recombination
cycle of atomic carbon. The reduced network consisting of less than 15 species
and about 20 reactions leads to less than 20\% uncertainty during the entire
computational time. An example of such a network for $R=100$~AU is given in
Table~\ref{m5_5}.

\begin{table}
\caption{Reduced network for UV-dominated chemistry in the surface layer.}
\begin{tabular}{lcl}
\hline
${\rm H}_2{\rm CO}^{\rm d}$&$\rightarrow$  &${\rm H}_2{\rm CO}$\\
${\rm C}^+  +  {\rm H}_2{\rm CO} $&$\rightarrow$  &$   {\rm H}_2{\rm CO}^+  +{\rm C}$\\
${\rm C}^+  +   e^-     $&$\rightarrow$  &$  {\rm C}  +   h\nu$\\
${\rm HCO}^+ +  e^-     $&$\rightarrow$  &$ {\rm CO}  +{\rm H}$\\
${\rm H}_2{\rm CO}^+ +  e^-     $&$\rightarrow$  &$  {\rm CO} + {\rm H} +{\rm H}$\\
${\rm H}_2{\rm CO}^+  + e^-     $&$\rightarrow$  &$  {\rm H}_2{\rm CO}  +h\nu$\\
${\rm C}   +    h\nu    $&$\rightarrow$  &$ {\rm C}^+  +  e^-$\\
${\rm CO} +     h\nu    $&$\rightarrow$  &${\rm C}+{\rm O}$\\
${\rm H}_2{\rm CO} +  h\nu    $&$\rightarrow$  &$  {\rm CO}  + {\rm H}_2$\\
${\rm H}_2{\rm CO} +h\nu    $&$\rightarrow$  &$  {\rm CO} + {\rm H}+{\rm H} $\\
${\rm H}_2{\rm CO} +h\nu    $&$\rightarrow$  &$  {\rm H}_2{\rm CO}^+ + e^-$\\
${\rm H}_2{\rm CO} +h\nu    $&$\rightarrow$  &$ {\rm HCO}^+  + e^- + {\rm H}$\\
\hline
\end{tabular}
\label{m5_5}
\end{table}

\section{Column densities}
\label{colden}

A useful product of our study is the complete chemical structure of the
disc. We present the calculated column densities of some species along with
observed values and column densities obtained by other groups in
Table~\ref{colden370}. The observed values are taken from Aikawa et
al.~(\cite{warm}), Table~1. Five theoretical models of the chemical evolution
of a protoplanetary disc are considered. The main differences between
them are discussed below. 

Willacy \& Langer~(\cite{wl00}) used the gas-phase
UMIST\,95 database, supplied with an additional set of gas-grain
processes and surface reactions. They adopted an extrapolated flared T~Tau
disc model of Chiang \& Goldreich~(\cite{CG97}), taking into account cosmic rays
and stellar UV radiation and assuming a sticking probability $S=1$.
In all the other models, namely, in Aikawa \& Herbst~(\cite{ah1999,ah2001}),
Aikawa et al.~(\cite{warm}), and van Zadelhoff et al.~(\cite{vzdetal}), the
New Standard Model (NSM) with a set of gas-grain interaction
processes is used as the chemical network.

Aikawa \& Herbst~(\cite{ah1999,ah2001}) considered an extrapolated
minimum-mass solar nebula model of Hayashi~(\cite{mmsn}), taking into account
stellar X-ray radiation in addition to the UV radiation and cosmic rays and assuming
an artificially low sticking probability $S=0.03$ that compensates for the absence
of non-thermal desorption mechanisms. The flared T~Tau disc model of D'Alessio et
al.~(\cite{DALdisc}) for various accretion rates is used with a sticking
probability $S=1.0$ in
Aikawa et al.~(\cite{warm}) and van Zadelhoff et al.~(\cite{vzdetal}).
Stellar X-rays are not included in the latter two models.

\begin{table*}
\caption{The observed and calculated column densities in cm$^{-2}$, $r=370$~AU, $t=10^6$~yrs.}
\label{colden370}
\tabcolsep 1mm
\begin{tabular}{llllllllllll}
\hline
\hline
Species   &  I      &     II   &   III    &    IV    &     V      &\multicolumn{2}{c}{This paper}& \multicolumn{3}{c}{Observed} & Max. \\
          &         &          &          &          &            &\multicolumn{2}{c}{$\dot{M}$, $M_\odot$ yr$^{-1}$}& DM Tau         & LkCa15$^*$    & LkCa15$^{**}$ & difference \\
          &         &          &          &          &            &  $10^{-9}$  &$10^{-7}$& (SD)           & (IM)          & (SD) & \\
\hline
H$_2$     &  9(22)  &  9(21)   & $1.5(24)$& $1.3(22)$& 1.3(23)    & $4.8(21)$  & $1.4(24)$& $3.8(21)$   &    ---        &    ---     & \\
\hline
H$_2$O    & ---     &  ---     &  ---     &  2.7(14) & 8.0(13)    & $2.1(14)$  & $1.6(14)$&   ---       &    ---        &    ---     & $\sim 3$ \\
HCN       & 2(12)   &  2(12)   & 2.1(12)  &  1.4(12) & 2.5(12)    & $1.3(13)$  & $6.7(12)$& $2.1(12)$   & $0.02-1.2(15)$&  $7.8(13)$ & $\sim 10$ \\
NH$_3$    & ---   & 1(13)   &  ---        &  1.4(13) & ---        & $4.4(13)$  & $1.4(14)$&   ---       &    ---        &    ---     & $\sim 15$ \\
\hline
C$_2$H    & 2(12)   &  8(12)   & 6.2(12)  &  8.9(11) & 2.0(13)    & $7.0(12)$  & $1.1(13)$& $4.2(13)$   &    ---        &    ---     & $\sim 20$ (10)\\
HNC       & 1(12)   &  ---     &  2.0(12) &  3.0(11) & ---        & $6.9(12)$  & $8.0(12)$& $9.1(11)$   & $<5.4(12)$    &    ---     & $\sim 30$  (8)\\
CN        & 1(13) & 1(13) &  3.8(12)      &  5.3(11) & 3.0(13)    & $6.4(12)$  & $4.6(12)$& $9.5-12(12)$& $9.1-25(13)$  &  $6.3(14)$ & $\sim60$ (8)\\
N$_2$H$^+$& ---     &  ---     &  1.9(12) &  8.1(11) & ---        & $6.0(12)$  & $8.1(13)$& $<7.6(11)$  & $<5.7(12)$    &  $<5.9(13)$& $\sim100$ (40)\\
HCO$^+$   & 3(11) & 5(11)  & $9.0(12)$    &  8.8(10) & 9.0(12)    & $8.4(11)$  & $2.0(11)$& $4.6-28(11)$& $1.5(13)$     &  $1.4(13)$ & $\sim 100$ (45)\\
CS        & 3(11) & 1(12)  & $4.9(11)$    &  1.0(11) & 1.2(12)    & $5.6(12)$  & $1.1(13)$& $6.5-13(11)$& $1.9-2.1(13)$ &  $2.2(14)$ & $\sim110$ (35)\\
\hline
CO        & 2(15) & 8(15)  & $1.1(18)$    &  7.1(16) & 1.2(18)    & $1.6(16)$  & $2.5(16)$& $5.7(16)$   & $1.6(18)$     &  $9.0(17)$ & $\sim 600$\\
H$_2$CO   & 7(11) & 1(12)  & $2.9(12)$    &  8.7(10) & ---        & $8.0(13)$  & $8.3(13)$& $7.6-19(11)$&    ---        &$3.0-22(13)$& $\sim10^3$\\
CH$_3$OH  & ---   &  ---    & $6.4(08)$   &  7.7(11) & ---        & $4.0(10)$  & $2.9(10)$&   ---       & $7.3-18(14)$  &  $<9.4(14)$& $<10^3$\\
\hline
\end{tabular}

   \begin{list}{}{}
        \item[I -- ] Aikawa \& Herbst (1999), Fig.~7, $R=400$~AU, $t=9.5\cdot10^5$~yrs, high $\zeta$ case
        \item[II -- ]  Aikawa \& Herbst (2001), Fig.~6, $R=400$~AU, $t=9.5\cdot10^5$~yrs
        \item[III -- ]  Aikawa et al. (\cite{warm}), Table~1, $\dot{M}=10^{-7}\,M_\odot$ yr$^{-1}$, $R=370$~AU, $t=9.5\cdot10^5$~yrs
        \item[IV -- ] Willacy \& Langer (2000), Table~4, interpolated to $R=400$~AU, $t=10^6$~yrs
        \item[V -- ] van Zadelhoff et al. (\cite{vzdetal}), Fig.~5, $R=400$~AU
        \item[IM -- ] interferometric observations, beam size is $\sim 4\arcsec$
        \item[SD -- ] single-dish observations, beam size is $\sim 20\arcsec$
        \item[$^*$ -- ] do not necessarily correspond to $r=370$~AU,
        \item[$^{**}$ -- ] disc size is $\sim 100$~AU
   \end{list}

\end{table*}

Column densities of selected species presented in these papers
are compared to those computed in this work in Table~\ref{colden370}.
Wherever possible, we took values from tables, otherwise figures were
used. Throughout this paper, a disc model with an accretion rate of
$10^{-7}\,M_\odot$ yr$^{-1}$ is used, but in Table~\ref{colden370} we
also present results of time-dependent chemical models
for a lower accretion rate, $\dot{M}=10^{-9}\,M_\odot$ yr$^{-1}$.

To characterise differences between model predictions, we divide all
species in three groups by the maximum ratio of their theoretical
column densities (less than 20, less than 200, and the rest). Given the
wide variety of conditions and assumptions
made in the various models, it is natural to expect that
differences in calculated column densities should be significant.
However, in reality, column densities for many species are close to each other
in the different studies. For example, the scatter in HCN column densities does not exceed an order
of magnitude. Similarly, ammonia densities in all but one case are a few
times $10^{13}$~cm$^{-2}$. Note that we omitted the anomalously high NH$_3$ column density
computed by Aikawa \& Herbst (\cite{ah1999}) as this was based on an
unrealistic rate for the H$_3^++\mathrm{N}$ reaction (Aikawa \& Herbst~\cite{ah2001}).

Somewhat higher maximum-to-minimum ratios for species in the second group
are mainly caused by their low abundances in the Willacy \& Langer~(\cite{wl00}) model.
In the last column of Table~\ref{colden370} we give relative variations
of their column densities both with and without (in parentheses) Willacy
\& Langer~(\cite{wl00}) data. If the latter are not taken into account, the
ratio of maximum to minimum column density does not exceed 50 for all these
molecules. This similarity of column densities and a lack of correlation between them
and the column density of molecular hydrogen can be understood as a further
manifestation of the ``warm molecular layer'' (Aikawa et al.~\cite{warm}) where nearly
all admixture gas-phase molecules are concentrated. On the other hand, molecular hydrogen
is concentrated in the midplane, where all other molecules are frozen out and
thus do not contribute to column densities.

The reason why column densities in the calculations of Willacy \& Langer are lower in
comparison with other studies is probably related to the disc model they adopted.
In the model of Chiang \& Goldreich~(\cite{CG97}) the disc is assumed to consist
of only two layers, namely, the dark dense 
midplane and the less dense surface layer subject to harsh UV radiation. 
In the cold midplane molecules are mainly locked in icy mantles while in the surface layer
they are easily dissociated or ionised by UV photons. In contrast to the disc model of 
D'Alessio et al.~(\cite{DALdisc}), there is no region similar to the shielded intermediate 
(``warm molecular'') layer, which is the most favourable disc domain for many molecules to have their 
maximum gas-phase concentrations.

The most striking finding is variations in theoretical CO column densities. As carbon monoxide, like other 
molecules, has its highest 
abundance in the intermediate layer, we might expect that its abundance is relatively independent of the 
total gas column density. Thus, the origin of the large discrepancies in CO abundances should lie in 
physical differences between various models. Looking at Table~\ref{colden370}, one is tempted
to assume that the CO abundance is low in models with surface chemistry (this paper and Willacy \& Langer~\cite{wl00})
or in models with X-ray induced chemistry (this paper and Aikawa \& Herbst~\cite{ah1999,ah2001}).
In the former case the CO molecule might be transformed to CO$_2$ and H$_2$CO, in the latter
case it might be destroyed by abundant He$^+$.

To check if this is the case, we computed the vertical distribution of CO for $R=370$~AU without surface reactions 
and with $\zeta_\mathrm{X}$=0. In both cases we failed to reproduce the high CO column densities
obtained by Aikawa et al. (\cite{warm}) and van Zadelhoff et al. (\cite{vzdetal}).
As we noted above, the low CO column density in the Willacy \& Langer model can partly be caused by
the disc model they adopted.
A possible explanation for the low CO abundance in our model is the fact that we do not take into account 
self- and mutual-shielding of molecular hydrogen and carbon monoxide, contrary to Aikawa et 
al.~(\cite{warm}) and Zadelhoff et al.~(\cite{vzdetal}). 

In addition, we include in Table~\ref{colden370} observationally inferred column densities
for DM~Tau (single-dish measurements) and LkCa15 (interferometric data). Note that
DM Tau values correspond to column densities averaged over the entire disc
($\sim 800$~AU), while values estimated from LkCa15 single-dish 
observations are column densities averaged over a $\sim 100$~AU disc.
For species in the first and second groups there is a reasonable agreement between
our predictions and observational data.

\section{Discussion}
\label{diss}

Recent theoretical studies (e.g. Aikawa \& Herbst~\cite{ah1999,ah2001}, Aikawa et 
al.~\cite{warm}, van Zadelhoff et al.~\cite{vzdetal}) revealed that the distribution
of gas-phase molecular abundances within a steady-state accretion disc in the absence
of mixing processes has a three-layer vertical structure.
In the dense and cold midplane all species, except the most volatile ones, are frozen out on grain surfaces
while in the disc atmosphere (surface layer) they are easily destroyed by high-energy stellar radiation. 
Therefore, in these disc domains gas-phase abundances of polyatomic species are low. On the contrary, in 
the intermediate layer, which is shielded from UV photons but still
warm enough to allow effective desorption, most molecules remain in the gas phase and drive a complex 
chemistry resulting in a rich molecular composition. As we noted above,
this simple picture may not be appropriate when diffusion processes are taken into account (Ilgner et al.~\cite{IHMM03}).

\begin{figure}
\begin{center}
\includegraphics[width=0.47\textwidth,clip]{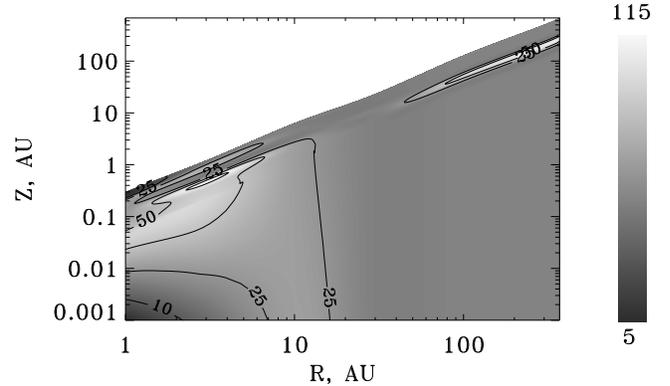}
\end{center}
\caption{The distribution of the number of species in the reduced networks governing fractional
ionisation over the disc. The highest and smallest number of species is 112 and 5, respectively.}
\label{rne_disc}
\end{figure}

In our study we focused on the analysis of chemical processes relevant to the evolution of
disc fractional ionisation based on the reduction approach. The size of a reduced chemical network,
accurately reproducing the electron concentration in a given disc region, can be understood as a quantitative
criterion for the complexity of ionisation chemistry. In Figure~\ref{rne_disc} we show
the distribution of the number of species in the reduced networks over the disc. This distribution
demonstrates a layered pattern. In most parts of the
disc the chemistry of ionisation is simple either due to the lack of ionising factors,
low temperature and high density (midplane) or due to the presence of ionising radiation (surface layer). 
In these regions, it is sufficient to keep about 10--30 species and a few tens of reactions in reduced networks.
Between the midplane and the surface layer the intermediate layer is located, where the evolution of ionisation 
degree is more complicated from the
chemical point of view (especially at early times). Here, one has to retain more than fifty species and a  
comparable number of reactions in the reduced networks. 

This layered structure is directly related to the size and location of the so-called ``dead zone''
within a protoplanetary disc, which is the region where the ionisation degree is so low that the
magnetic field is not coupled to the gas. The MHD turbulence
cannot develop in this region. Thus, the MRI is not operative, which means the lack of an effective angular momentum 
transport if no other transport mechanism can be identified. The ``dead zone'' in accretion discs has
been widely investigated (e.g. Igea \& Glassgold~~\cite{IG99}, Fromang et al.~\cite{from2002},
Sano et al.~\cite{sano}). 

An important quantity which characterises the coupling between the matter and the magnetic field is
the magnetic Reynolds number, which can be defined as (Fromang et al.~\cite{from2002})
\begin{equation}
\label{rem}
  Re_M = \frac{c_sH}{\eta}, 
\end{equation}
where $\eta$ is the Ohmic resistivity (Blaes \& Balbus~\cite{bb1994})
\begin{equation}
\label{eta}
  \eta = \frac{234\cdot T^{0.5}}{x_\mathrm{e}}.
\end{equation}
The other quantities are the sound speed $c_s$, and the thickness of the disc $H$.
The MHD turbulence can be maintained only if this quantity exceeds a certain critical value which
depends on the field geometry and other factors. Following Fromang et al., we consider two cases of this limiting
value, namely, $Re_M^\mathrm{crit}=100$ and $Re_M^\mathrm{crit}=1$ and define the ``dead zone'' as a
disc domain where $Re_M<Re_M^\mathrm{crit}$. 

\begin{figure}
\begin{center}
\includegraphics[width=0.47\textwidth,clip]{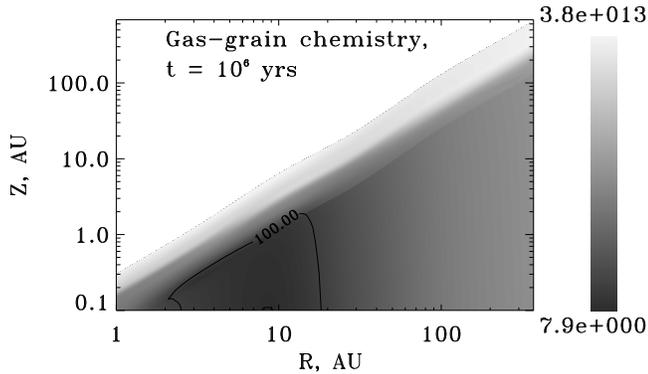}
\end{center}
\caption{The magnetic Reynolds numbers computed for our model and an evolutionary time of $10^6$~yrs. 
The solid line designates $Re_M^\mathrm{crit}=100$. The highest and lowest values are equal to
about $10^{13}$ and 10, respectively.}
\label{re_m_10_6}
\end{figure}

The calculated magnetic Reynolds numbers for our model are shown in Fig.~\ref{re_m_10_6}. 
The lowest magnetic Reynolds number is $Re_M \sim 10$,
which implies that under certain circumstances within the framework of our model the ``dead zone'' is entirely absent.
With another critical value, $Re_M^\mathrm{crit}=100$, the ``dead zone'' occupies the following disc
region: $2~\mathrm{AU}< r < 20$~AU, $z\sim 0.06\cdot r^{1.27}$ (solid line). This result is roughly
consistent with other recent studies. For example, Fromang et al. found that the ``dead zone'' can
extend over $0.7~\mathrm{AU}< r < 100$~AU, if the viscosity parameter is equal to $10^{-2}$ and the accretion rate
is $10^{-7}\,M_\odot$~yr$^{-1}$, whereas Igea \& Glassgold~(\cite{IG99}) obtained a ``dead zone'' of somewhat 
smaller size, $r< 5$~AU.

Apart from defining the location of the ``dead zone'' in accretion discs,
the fractional ionisation determines whether non-ideal MHD effects are
important for the evolution of protoplanetary discs. When 
the electron abundance is computed for the solution of non-ideal
MHD equations, an ionisation equilibrium 
is often assumed. An expression similar to Eq.~(\ref{xe}) is then used to calculate the equilibrium
fractional ionisation, with different estimates for the 
gas-phase recombination coefficient $\beta$ (Blaes \& Balbus~\cite{bb1994}; Gammie~\cite{gammie};
Regos~\cite{regos}; Reyes-Ruiz~\cite{rr2001}; Fromang et al.~\cite{from2002};
Fleming \& Stone~\cite{fs2003}).

\begin{figure}
\begin{center}
\includegraphics[width=0.45\textwidth,clip]{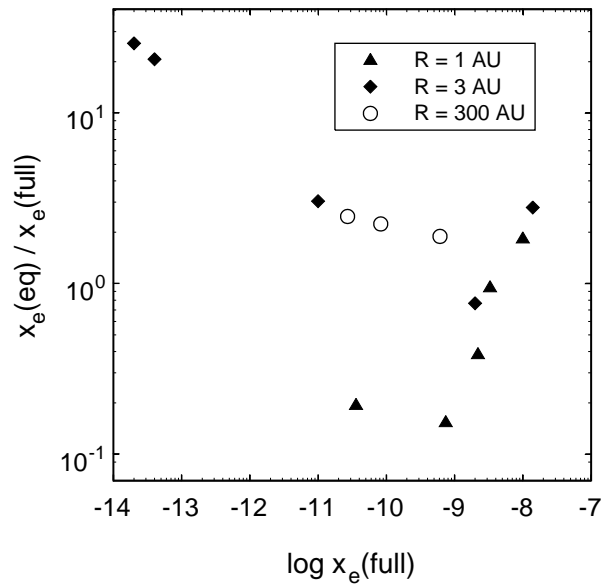}
\end{center}
\caption{Comparison of the equilibrium and time-dependent fractional ionisations
at $t=10^6$~years in the midplane and in the intermediate layer. Symbols correspond to different
radii.}
\label{eqvsfull}
\end{figure}

In Fig.~\ref{eqvsfull} we compare the fractional
ionisation $x_\mathrm{e}(\mathrm{eq})$, computed from Eq.~(\ref{xe}) with the often used
expression for recombination coefficient $\beta=8.7\cdot10^{-6}T^{-1/2}$
(Glassgold, Lucas, \& Omont~\cite{beta1986}), to the fractional ionisation computed
with the full chemical network $x_\mathrm{e}(\mathrm{full})$, for some representative
points in the midplane and in the intermediate layer. It is not surprising
that the ratio of these two quantities exceeds 10 at low ionisation degrees.
Here the electron density is small and ion-grain interactions should be taken into account.
At moderate ionisation degrees ($10^{-10}-10^{-8}$), the equilibrium
value may differ from the ``exact'' value by a factor of a few
at the end of the computation. At earlier times, the discrepancy is higher.

However,
in the midplane the equilibrium state is reached very rapidly, with no appreciable changes
in $x_\mathrm{e}$ after $\sim1000$~years of evolution. If one is
not interested in time scales less than 1000~years and can afford a factor of a few
uncertainty in the fractional ionisation, deep in the disc interior
the equilibrium $x_\mathrm{e}$ is sufficient, provided
grain charging processes are taken into account.

The intermediate layer shows a very different behaviour. As we mentioned previously, the equilibrium
state is not reached there or is reached very late. The initial fall-off of $x_\mathrm{e}$, caused by HCO$^+$,
H$_3^+$ and metal ion recombination, is followed by a step-like increase. The duration
of each phase and $x_\mathrm{e}$ evolutionary changes, illustrated in Figure~\ref{fmed24},
vary significantly in different parts of the layer. Surprisingly, the equilibrium
$x_\mathrm{e}$ from Eq.~(\ref{xe}) is very close to the computed fractional ionisation
at $t=10^6$~years in this point ($1.5\cdot10^{-9}$ and $2\cdot10^{-9}$, respectively).
This is, of course, a coincidence, as the equilibrium is not reached there.
We emphasise that even if $x_\mathrm{e}$(full) is equilibrated eventually
in the intermediate layer and is close to $x_\mathrm{e}$(eq), the latter value
does not reflect the ionisation state of the medium for most of the computational time.

In the surface layer the fractional ionisation is controlled not only by X-rays, but also
by UV-radiation. The equation (\ref{xe}) is still applicable here but with an important
change. In the surface layer $x_\mathrm{e}$ is
determined mostly by ionised carbon (Table~\ref{domions}). Its recombination coefficient is much
lower than the value quoted above, thus, $\beta=8.7\cdot10^{-6}T^{-1/2}$
should not be used in this case. If one substitutes in Eq.~(\ref{xe}) the correct $\beta$
for C$^+$ ($1.4\cdot10^{-13}T^{-0.61}$), the resultant equilibrium $x_\mathrm{e}$ is
a few times smaller than the ``exact'' value, mainly because photoionisation is not
taken into account. Time needed to reach the equilibrium $x_\mathrm{e}$ in the
surface layer does not exceed 100~years. It is questionable whether precise information
about such a high ionisation ($x_\mathrm{e}\sim10^{-4}$) is required for
MHD computation.

Overall, we conclude that the equilibrium approach is appropriate most of the time
and in most parts of the studied disc, but it should be applied with care. This is especially
important in less dense discs that are more transparent to X-rays and UV radiation.
In this paper, we used a relatively massive disc model with a narrow intermediate
layer. In models with lower accretion rates the intermediate layer would occupy a greater
volume of the disc.

Even though the reduced networks presented in this paper provide
much better accuracy than the equilibrium approach, while having
quite modest sizes, their usefulness for dynamical models is limited
by the fact that they are computed in a steady-state disc.
For example, in the point M1 the reduced network consists
of three species, NO, Mg, and Mg$^+$. This network is of no use in
a dynamical model because there will be no NO in the gas flowing
into the inner part of the midplane, as reduced networks further
out in the midplane do not contain this molecule. To make the
reduction valuable for MHD modelling we need to consider
it in a dynamically evolving medium.

An alternative approach is to merge all the networks described in this paper into a
single network that comprises about 120 species and a comparable number of reactions.
This network is out of the scope of the present paper and will be described in
further publications.

\section{Conclusion}
\label{concl}

Chemical processes, responsible for the ionisation structure of a protoplanetary
disc with a central star, are analysed by means of reduced
networks that reproduce the ionisation fraction within a factor of 2. These
networks are available from the authors upon request. Because of the wide range of physical
conditions met in a typical disc, there is a corresponding diversity in the chemical
reactions that control the fractional ionisation in different parts of the
disc. Generally, it can be divided into three layers. In the midplane the ionisation
is provided by cosmic rays and radioactive elements only. Above the midplane, the
intermediate layer is located where the ionisation is dominated by X-rays. In the surface 
layer UV photons are the main ionising factor. In each of these layers we analyse several representative
points and construct reduced chemical networks that are needed to reproduce the fractional
ionisation as a function of time during $10^6$ years of evolution within a factor of
2~uncertainty. In the midplane the chemistry, which determines the fractional ionisation,
is very simple. Reduced networks typically include no more than 10 species and
a similar number of reactions. The same holds true for the surface layer. The
intermediate layer is the most complicated region from the chemical point of view. To compute
the fractional ionisation with the targeted accuracy, in some regions one has to take into
account carbon chains containing up to 6~carbon atoms, which leads to
reduced networks with over a hundred species and reactions.

\begin{acknowledgements}
We are grateful to our referee, Dr. A. Markwick, whose comments helped to improve
and clarify our presentation significantly.
DS was supported by the \emph{Deut\-sche For\-schungs\-ge\-mein\-schaft, DFG\/} project 
``Research Group Laboratory Astrophysics'' (He 1935/17-2). DW acknowledges support by
the RFBR grants 01-02-16206 and 02-02-04008 and by the President of the RF grant MK-348.2003.02.
\end{acknowledgements}


\begin{thebibliography}{}

\bibitem[1999]{ah1999}
Aikawa, Y., \& Herbst, E. 1999, \aap, 351, 233

\bibitem[2001]{ah2001}
Aikawa, Y., \& Herbst, E. 2001, \aap, 371, 1107

\bibitem[2002]{warm}
Aikawa, Y., van Zadelhoff, G. J., van Dishoeck, E. F., Herbst, E. 2002, \aap, 386, 622

\bibitem[1991]{MRI}
Balbus, S. A., \& Hawley, J. F. 1991, \apj, 376, 214

\bibitem[1997]{bauer}
Bauer, I., Finocchi, F., Duschl, W. J., Gail, H.-P., Schloeder, J. P. 1997, \aap, 317, 273

\bibitem[1994]{bb1994}
Blaes, O. M., \& Balbus, S. A. 1994, \apj, 421, 163

\bibitem[1983]{BH1983}
Burke, J. R., \& Hollenbach, D. 1983, \apj, 265, 223

\bibitem[1997]{CG97}
Chiang, E., \& Goldreich, P. 1997, \apj, 490, 368

\bibitem[1999]{DALdisc}
D'Alessio, P., Calvet, N., Hartmann, L., Lizano, S., Canto, J. 1999, \apj, 527, 893

\bibitem[1978]{Draine}
Draine, B. T. 1978 \apjs, 36, 595

\bibitem[2003]{fs2003}
Fleming, T., \& Stone, J. 2003, \apj, 585, 908

\bibitem[2002]{from2002} 
Fromang, S., Terquem, C., Balbus, S. A. 2002, \mnras, 329, 18

\bibitem[1996]{gammie}
Gammie, Ch. 1996, \apj, 457, 355

\bibitem[1986]{beta1986}
Glassgold, A. E., Lucas, R., \& Omont, A. 1986, \aap, 157, 35

\bibitem[1997a]{zetaxa}
Glassgold, A. E., Najita, J., Igea, J. 1997a, \apj, 480, 344

\bibitem[1997b]{zetaxb}
Glassgold, A. E., Najita, J., Igea, J. 1997b, \apj, 485, 920

\bibitem[1992]{hhl92}
Hasegawa, T. I., Herbst, E., \& Leung, C. M. 1992, \apjs, 82, 167

\bibitem[1993]{hh93}
Hasegawa, T. I., \& Herbst, E. 1993, \mnras, 263, 589

\bibitem[1981]{mmsn}
Hayashi, C. 1981, Prog. Theor. Phys., 70, 35

\bibitem[1999]{IG99}
Igea, J., \& Glassgold, A. E. 1999 \apj, 518, 848

\bibitem[2003]{IHMM03}
Ilgner, M., Henning, Th., Markwick, A., Millar, T. 2003, \aap, in press

\bibitem[1997]{h2c6} 
Langer, W. D., Velusamy, T., Kuiper, T. B. H., Peng, R., McCarthy, M. C.,
et al. 1997, \apjl, 480, L63

\bibitem[1985]{leger}
L\'eger, A., Jura, M., Omont, A. 1985, \aap, 144, 147

\bibitem[2002]{markwick}
Markwick, A. J., Ilgner, M., Millar, T. J., \& Henning, Th. 2002, A\&A, 385, 632

\bibitem[1997]{umist95}
Millar, T. J., Farquhar, P. R. A., \& Willacy, K. 1997, A\&AS, 121, 139

\bibitem[1991]{Nishi}
Nishi, R., Nakano, T., Umebayashi, T. 1991, \apj, 369, 181

\bibitem[1992]{ohishi}
Ohishi, M., Irvine, W. M., \& Kaifu, N. 1992, in Astrochemistry of Cosmic
Phenomena, ed. P. D. Singh (Dordrecht: Kluwer), 171

\bibitem[1974]{OD74}
Oppenheimer, M., \& Dalgarno, A. 1974, \apj, 192, 29
 
\bibitem[2002]{rh2002}
Roberts, H., Herbst, E. 2002, \aap, 395, 233

\bibitem[1997]{regos} 
Regos, E. 1997, \mnras, 286, 97

\bibitem[2001]{rr2001} 
Reyes-Ruiz, M. 2001, \apj, 547, 465

\bibitem[2002]{Rea02} 
Ruffle, D. P., Rae, J. G. L., Pilling, M. J., Hartquist, T. W., \& Herbst, E.
2002, \aap, 381, L13

\bibitem[2000]{sano}
Sano, T., Miyama, Sh. M., Umebayashi, T., \& Nakano, T. 2000, \apj, 543, 486

\bibitem[1998a]{th1998a}
Terzieva, R., Herbst, E. 1998a, \apj, 501, 207

\bibitem[1998b]{th1998b}
Terzieva, R., Herbst, E. 1998b, \apj, 509, 932

\bibitem[1980]{un1980}
Umebayashi, T., \& Nakano, T.  1980, \pasj, 32, 405

\bibitem[2003]{vzdetal}
van Zadelhoff, G.-J., Aikawa, Y., Hogerheijde, M. R., \& van Dishoeck E. F.
2003, \aap, 397, 789

\bibitem[1999]{wpf1999}
Walmsley, C. M., Pineau des For\^ ets, G., \& Flower, D. R. 1999, \aap, 342, 542

\bibitem[1995]{westley}
Westley, M. S., Baragiola, R. A., Johnson, R. E., Baratta, G. A. 1995, Nature, 373, 405

\bibitem[2003]{papi}
Wiebe, D., Semenov, D., \& Henning, Th. 2003, \aap, 399, 197 (Paper~I)

\bibitem[1998]{willacy}
Willacy, K., Klahr, H. H., Millar, T. J., Henning Th. 1998, \aap, 338, 995

\bibitem[2000]{wl00}
Willacy, K., Langer, W. 2000, \aap, 397, 789

\end{thebibliography}
\end{document}